\documentclass[10pt, conference, compsocconf]{IEEEtran}
\IEEEoverridecommandlockouts
\usepackage{amsmath}
\usepackage{amssymb}
\usepackage{amsfonts}

\usepackage{algorithmic}
\usepackage[ruled,linesnumbered]{algorithm2e}
\usepackage{siunitx} 

\usepackage{graphicx}
\usepackage{xcolor}
\usepackage{textcomp}
\usepackage{epsfig}

\usepackage{comment} 

\usepackage{url}
\usepackage[bookmarks=false]{hyperref}

\usepackage{microtype}
\usepackage{booktabs} % for professional tables

\usepackage{times}
\usepackage{shortcuts_js}

\usepackage{cleveref} 

\def\BibTeX{{\rm B\kern-.05em{\sc i\kern-.025em b}\kern-.08em
    			 T\kern-.1667em\lower.7ex\hbox{E}\kern-.125emX}}
    			 
\newcommand{\showComments}{yes}

\newcommand{\note}[2]{
    \ifthenelse{\equal{\showComments}{yes}}{\textcolor{#1}{#2}}{}
}

\newcommand{\name}{Polystore++\xspace}

\begin{document}

\title{Polystore++: Accelerated Polystore System for Heterogeneous Workloads\vspace{-15pt}}

\author{
\IEEEauthorblockN{Rekha Singhal\textsuperscript{*}, Nathan Zhang, Luigi Nardi, Muhammad Shahbaz, and Kunle Olukotun 
\thanks{\noindent\textsuperscript{*}
Rekha Singhal is a Visiting Scholar at Stanford University and a Senior Scientist at Tata Consultancy Services, India. Email:~\url{rekha.singhal@tcs.com}.}
}
\IEEEauthorblockA{\textit{Department of Computer Science} \\
\textit{Stanford University} \\
\{rekhas2,stanfurd,lnardi,shahbaz3,kunle\}@stanford.edu}
}  

%\author{
%\IEEEauthorblockN{Rekha Singhal\textsuperscript{*}
%\thanks{\noindent\textsuperscript{*}
%Rekha Singhal is a Visiting Scholar at Stanford University and a Senior Scientist at Tata Consultancy Services, India (email: \href{rekha.singhal@tcs.com}{rekha.singhal@tcs.com})}}
%\IEEEauthorblockA{\textit{Computer Science} \\
%\textit{Stanford} \\
%CA, USA \\
%rekhas2@stanford.edu}
%\and
%\IEEEauthorblockN{Nathan Zhang}
%\IEEEauthorblockA{\textit{Computer Science} \\
%\textit{Stanford} \\
%CA, USA \\
%stanfurd@stanford.edu}
%\and
%\IEEEauthorblockN{Luigi Nardi}
%\IEEEauthorblockA{\textit{Computer Science} \\
%\textit{Stanford} \\
%CA, USA \\
%lnardi@stanford.edu}
%\and
%\IEEEauthorblockN{Kunle Olukotun}
%\IEEEauthorblockA{\textit{Computer Science} \\
%\textit{Stanford} \\
%CA, USA \\
%kunle@stanford.edu}
%}

\maketitle
\thispagestyle{plain}

\begin{abstract}

Modern real-time business analytic consist of heterogeneous workloads (\eg database queries, graph processing, and machine learning). These analytic applications need programming environments that can capture all aspects of the constituent workloads (including data models they work on and movement of data across processing engines). Polystore systems suit such applications; however, these systems currently execute on CPUs and the slowdown of Moore's Law means they cannot meet the performance and efficiency requirements of modern workloads. We envision \name{}, an architecture to accelerate existing polystore  systems using hardware accelerators (\eg FPGAs, CGRAs, and GPUs). \name{} systems can achieve high performance at low power by identifying and offloading components of a polystore system that are amenable to acceleration using specialized hardware. Building a \name{} system is challenging and introduces new research problems motivated by the use of hardware accelerators (\eg optimizing and mapping query plans across heterogeneous computing units and exploiting hardware pipelining and parallelism to improve performance). In this paper, we discuss these challenges in detail and list possible approaches to address these problems. 

\end{abstract}

\section{Introduction}
\label{sec:introduction}

\begin{figure}[t]
  \centering
  \includegraphics[width=0.9\linewidth]{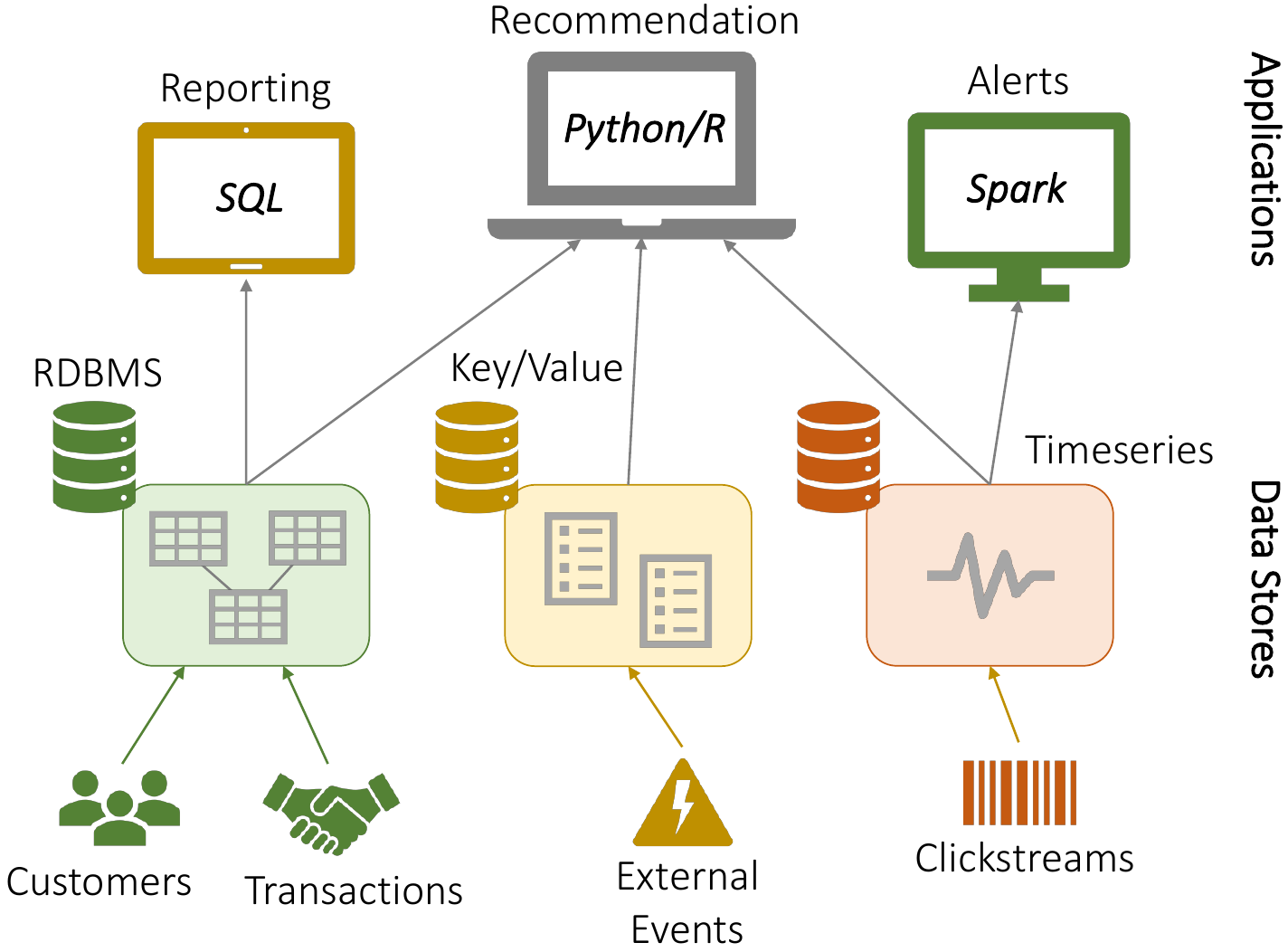}
  \caption{Examples of enterprise analytic applications. A recommendation system spans across multiple heterogeneous data stores (\ie RDBMS, Key/Value, and Timeseries).}
    \label{fig:polystore-eg1}
\end{figure}

\begin{figure}[t]
  \centering
  \includegraphics[width=\linewidth]{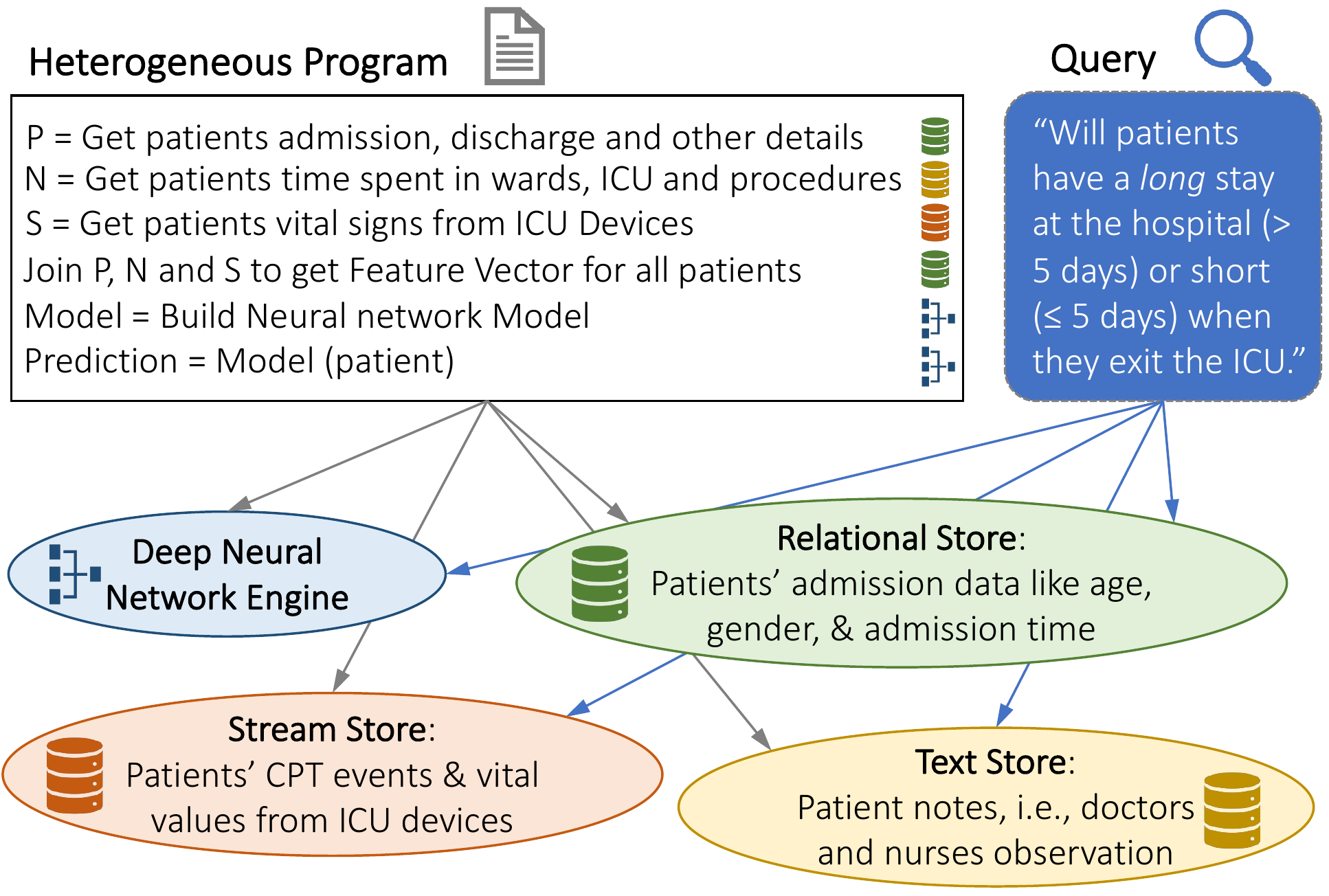}
  \caption{A real-time application for clinical analysis with an example of natural-language query and the corresponding heterogeneous program for the MIMIC III dataset~\cite{mimicapp}. Each step in the heterogeneous program is a workload that interacts with relational, text, and stream store, as well as a deep neural network engine.}
    \label{fig:polystore-eg2}
\end{figure}

Modern data-analytic applications such as personalized health care~\cite{mimicapp}, content filtering, and monitoring~\cite{content} often process data segregated across legacy data-processing engines (\eg Oracle, Neo4j, and Spark) each with its own custom data models; such applications are referred to as heterogeneous workloads~\cite{flare}. For example, enterprises typically maintain transactional records in a relational store (like Postgres~\cite{postgres}) and  users' clickstreams in a timeseries store (like TimescaleDB~\cite{timeseries}). A recommendation application may need to access both the transactional and clickstream data to predict the next best offers for users (Figure~\ref{fig:polystore-eg1}). As another example, consider the Mimic~III clinical dataset~\cite{mimicapp}: a comprehensive collection of information on patients admitted to the Beth Israel Deaconess Medical Center in Boston, MA, USA for the period between 2001 and 2012. It includes timeseries information from the bedside monitoring devices, waveforms from the history of previous patients, structured logs of patients' metadata (\eg patient address, phone number, and more in a relational table), as well as semi-structured data (\eg text) consisting of doctors' and nurses' notes and prescriptions. A real-time health-monitoring application for patients, in Figure~\ref{fig:polystore-eg2}, when operating on Mimic III would therefore rely on different data-processing engines for each of the given data types (\eg timeseries and text) to quickly predict patients' stay in ICU, typically with a latency target of a few milliseconds.
 
Typically, these analytic applications consist of a combination of multiple sub-applications, each written and optimized for a particular data-processing engine (\eg Oracle for SQL queries and Neo4j for graph-based operations). The number of such sub-applications depends on the different types of data models accessed by an analytic application. Each sub-application works on one of these data models, reading and processing the data using the corresponding data store. Finally, the processed data is moved across sub-applications (\eg in sequence), transforming it to the data model of the receiving application. Or, sub-applications submit their results to a central storage, mapping it to a single data model, which a post-processing application running on the associated engine processes to generate the final outcome---an approach commonly known as `one-size-fits-all.' However, both these approaches spend majority of the time performing unnecessary movement and remodeling of data, thus, inflating the response latency of a real-time analytic system. Furthermore, they burden the developers of the analytic applications to manually specify and optimize scheduling policies to achieve high performance.

A polystore system~\cite{bigdawg} on the other hand assumes that `one-size-does-not-fit-all,' and federates and automates processing of data across engines. In addition, these engines work with specific data models to execute heterogeneous workloads efficiently. Each processing engine (\eg a database system) is optimized for a specific set of operations, \eg joins in Postgres~\cite{postgres}, matrix operations in SciDB~\cite{scidb}, and path-finding in Neo4j~\cite{neo4j}. A polystore system exploits these domain-specific characteristics of these engines to expedite heterogeneous workloads. The system forms a complete stack consisting of frontend interfaces for application development, compilers and optimizers to generate efficient code, `CAST'~\cite{bigdawg} for moving data across various storage platforms, and native processing engines with their respective adapters.

Polystore systems minimize workload completion times by exploiting native capabilities of data-processing engines while optimizing for CPU-based computations. However, with Moore's Law~\cite{schaller1997moore} slowing down compute and Dennard Scaling~\cite{bohr200730} hitting power limits, we no longer observe exponential improvements in general-purpose compute using CPUs with fixed area and power cost. Therefore, industry is now moving toward domain-specific hardware architectures that trade flexibility of CPUs for improved efficiency. For example, Microsoft Brainwave~\cite{catapult} and Google's Tensor Processing Unit (TPU)~\cite{tpu} are designed to accelerate deep-learning workloads unlike general-purpose CPU based systems.

\begin{figure}[t]
  \centering
  \includegraphics[width=\linewidth]{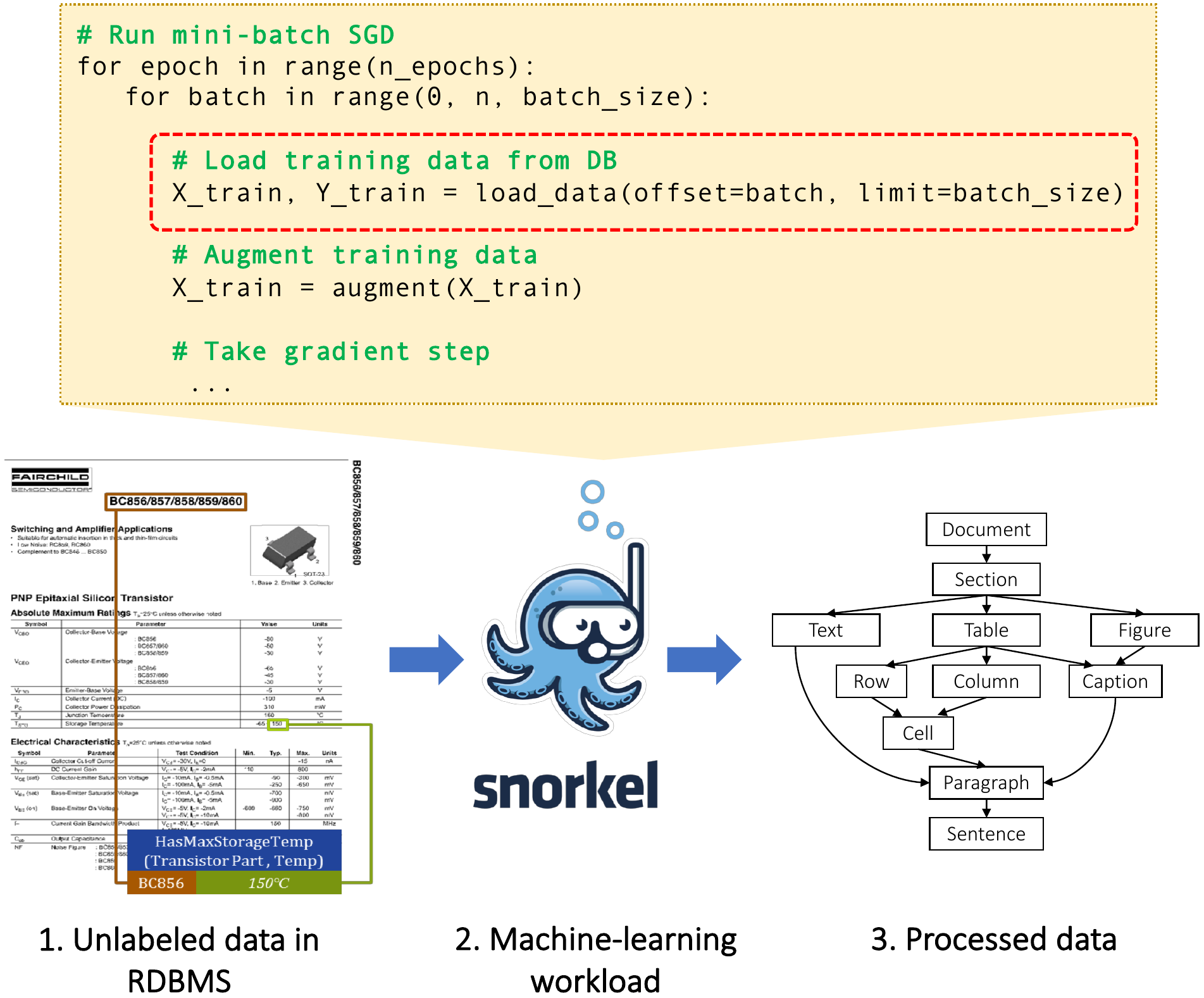}
  \caption{Snorkel: A deep-learning model pipeline using SQL calls in Python for labelling training data using weak supervision.}
  \label{fig:snorkel}
\end{figure}

We envision that next-generation polystore systems, or \name{} systems, can leverage such hardware accelerators~\cite{hwa} to reduce the execution time of heterogeneous workloads. For example, the Snorkel~\cite{snorkel} application (Figure~\ref{fig:snorkel}), is an example of a tight integration of SQL and machine-learning workloads.
The \verb|load_data| function is interspersed in ML code to fetch data from a database using SQL query. A polystore++ system can identify this mix and accelerate the \verb|load_data| function using hardware accelerators. 

\name{} systems deploy accelerators in standalone, coprocessor, or bump-in-the-wire modes~\cite{HWlandscape}. In standalone mode, key functions of a \name{} system run entirely on the hardware, whereas in the coprocessor mode the system logic is distributed across the host CPUs and hardware accelerators. And, bump-in-the-wire accelerators sit between the data-processing engines and data stores. However, these accelerators introduce new challenges of how to (1) facilitate development of clearly-specified (or clarity-optimized) heterogeneous programs, (2) convert heterogeneous programs into unified representations, (3) identify patterns in workloads to accelerate, (4) optimize execution plans across heterogeneous computing units (\eg CPU-based data-processing engines and hardware accelerators)---a multi-objective optimization, (5) generate optimized code specific to a computing target (\eg FPGA and GPU), and (6) schedule workloads to exploit hardware parallelism and pipelining. 

In this paper, we discuss how \name{} systems can achieve high performance at low power by identifying and offloading polystore system components that are amenable to acceleration using specialized hardware. We detail new research challenges that surface with the introduction of hardware accelerators in polystore systems and discuss possible approaches to address these challenges~\S\ref{sec:challenges}.
 
We organize the paper as follows:~\S\ref{sec:related} discusses the state-of-the-art in the areas of polystore systems and hardware accelerators.~\S\ref{sec:arch} presents an architecture for \name{} systems and discusses opportunities for hardware acceleration.~\S\ref{sec:challenges} highlights the new challenges and approaches introduced due to accelerators and heterogeneous programs in polystore systems ranging from workload optimization, scheduling, and execution in \name{} systems.

\section{Related Work}
\label{sec:related}

\subsection{Polystore Systems}
%Polystore systems have evolved from federated databases processing SQL workloads on autonomous relational databases to processing of heterogeneous workload federated on data-processing engines working with different data models. Polystore should not be confused with Data virtualization systems such as Denodo~\cite{denodo} that addresses only unification of data coming from different types of data sources.
%• A polyglot system hosts data using a collection ofhomogeneous data stores and exposes multiple queryinterfaces to the users.
%• A multistore system is able to manage data acrossheterogeneous data stores, while supporting a singlequery interface.
%• A polystore system enables query processing acrossheterogeneous data stores and supports multiple queryinterfaces.
Ran et al.~\cite{polystoresurvey} presents a taxonomy of systems characterized by data stores and query interfaces that they support for workload processing. (1) The federated database systems such as Multibase~\cite{multibase} execute workloads on a set of relational data stores using only an SQL query interface. However, with the advent of complex analytic applications working on different data models (\eg relational, key-value, graph, and tensor), these systems became obsolete. (2) Polyglot systems such as Spark~\cite{onesize}, Weld~\cite{weld} and MyriaX~\cite{myriax}, are examples of 'one-size-fits-all,' that support multiple query interfaces and process heterogeneous workloads on homogeneous data stores. For example, GRfusion~\cite{grfusion} is a system that processes multiple queries for graph and machine learning workloads. Furthermore, Polyglot systems reduce workload completion time by fusing operators~\cite{weld} and executing them on a specialized data-processing engine. (3) Multistores process workloads across heterogeneous data stores using a single query interface. For example, HadoopDB~\cite{hadoopdb} processes queries across both relational and Hadoop data stores. It limits users to specify workloads using a single SQL-based interface. (4) Polystore systems combine advantages of both polyglot and multistore systems by supporting multiple query interfaces to represent disparate workloads and their execution across heterogeneous data stores. They utilize the native operations of data-processing engines to execute workload faster compared to previous systems~\cite{bigdawgql}.

BigDawg~\cite{bigdawg} is one of the first implementations of a polystore system, having a SQL-like interface extended with BigDawg operators, such as `bdrel' and `bdarray' to fetch data from relational and array stores.
Myria~\cite{myria} is a cloud service with an algebra-based optimizer for efficient execution on data federated across heterogeneous data stores. Myria uses Pipegen~\cite{pipegen} to migrate data across data-processing engines to improve performance (\S\ref{sec:arch}). 
Rheem~\cite{rheem} is another polystore system that builds a cost model for each data processing operator by collecting their resource utilization from respective data-processing engines. Teradata Vantage~\cite{teradata} is a commercial polystore system that supports different paradigms for workload processing on native data-processing engines. It facilitates application development without burdening developers to manually schedule tasks for data movement. However, users still have to explicitly transfer data from legacy database systems to the Teradata Vantage environment.
Performance improvements in these polystore systems have been limited to CPU-based computations and cannot keep pace with increasingly demanding modern heterogenous workloads.

\subsection{Hardware Accelerators}
Compute acceleration is the use of specialized hardware to perform functions more efficiently than general-purpose CPU~\cite{hwa}. These accelerators exploit hardware parallelism and pipelining to achieve high performance and maintain low power by operating at lower clock frequency and adding hardware support for a variety of operations.

Specialized hardware can accelerate polystore systems by leveraging multicore CPUs, graphics processing unit (GPUs~\cite{gpu}), field programmable gate arrays (FPGAs~\cite{fpga}), application-specific integrated chips (ASICs~\cite{asics}), or coarse grain reconfigurable arrays (CGRAs~\cite{cgra}) such as Plasticine~\cite{plasticine}. Multicore CPUs consume more power per task with limited parallelism and inefficient data movement compared to well-matched applications running on accelerators. GPUs, on the other hand, perform wide SIMD computations using hundreds of cores. GPUs clock these cores at lower speeds compared to CPUs, but achieve higher throughput by running workloads across many such cores. GPUs primarily focus on operations that exploit SIMD parallelism, such as matrix multiplication, and more recently database applications. FPGAs compared to CPUs and GPUs consume less power and allows programming more generalized operations. FPGAs can be used as a flexible accelerators for high-cost operations within an execution plan and are well suited for operations on streaming data.
%, and applications can exploit pipeline execution for higher throughput. 
However, these FPGAs are hard to program and can only support limited number of complex operations. Developers program FPGAs using hardware description languages (\eg Verilog or VHDL), with lengthy compile times to synthesize applications to an FPGA configuration (\ie a network of look-up tables, LUTs) leading to large development overheads. CGRAs have short reconfiguration time as they are constructed using standard processing elements~\cite{cgra}.
ASICs are fixed-function devices with pre-configured logic (\ie data-processing operators). These devices cannot be reconfigured but achieve extremely high performance and efficiency for these operators. 

A hardware accelerator can sit in the access path of data-processing operators (as in Netezza~\cite{netezza}). Or, it can run as a coprocessor, installed on PCIe slots of a data-processing server alongside the CPU to accelerate frequently occurring operations. For example, GPU-based accelerators can speed up Spark workloads (\eg SparkGPU~\cite{sparkgpu} and Flare~\cite{flare}); FPGA-based devices can accelerate relational databases (\eg Postgres), hybrid workloads (\eg DANA: native SQL extended for ML operators~\cite{DBaccl}), and streaming workloads (\eg Saber~\cite{saber}). Furthermore, these devices can operate in standalone mode (like Tensor Processing Unit~\cite{tpu} or Microsoft Brainwave~\cite{catapult}). An exhaustive survey of workload acceleration using these devices is presented in \cite{DBSurvey}.

However, hardware acceleration tradeoffs general-purpose compute to achieve high performance at low power, thus making these devices difficult to program. Each accelerator is programmed using its hardware-specific low-level language (\eg Verilog), which requires a developer to have deep understanding of the underlying hardware. Application and hardware domain-specific languages (DSLs), such as Spatial~\cite{spatial}, Relay~\cite{relay}, and Delite~\cite{delite}, ease application development for specific accelerators by abstracting low-level abstractions into high-level primitives (\eg parallel patterns~\cite{plasticine}) that a developer is familiar with. For example, Halide~\cite{halide} and Tensorflow~\cite{tensorflow} are application-specific DSLs for image processing and deep neural network processing respectively. 
 
Furthermore, each hardware accelerator has a specific area and power profile. Its design determines the corresponding performance and power benefits. Altaf and Wood proposed LogCA~\cite{logca} to model the performance of these accelerators based on their design and interface to the host system using applications with different computational intensity\footnote{Computational Intensity is the amount of work done on a host/accelerator per byte of offloaded data.}. LogCA helps in designing accelerator architecture for memory access and compute bound operators (or kernels).

\section{Architecture}\label{sec:arch}

\begin{figure}[t]
  \centering
  \includegraphics[width=\linewidth]{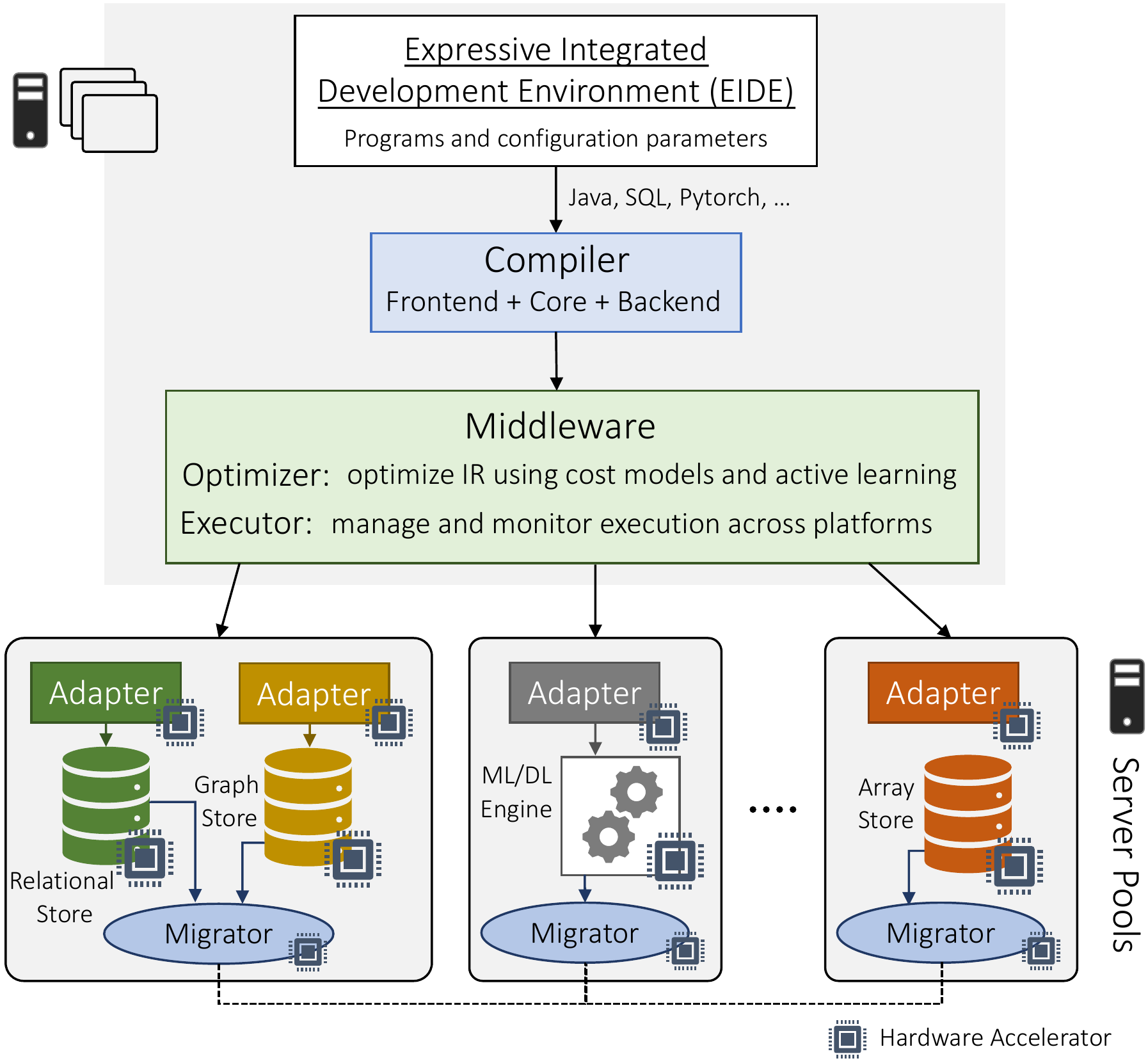}
  \caption{\name{} system's physical layout: hardware accelerators for data-processing engines, adapters, and data migrators running on a pool of servers; a middleware that parses, optimizes, and manages execution of heterogenous workloads; and an interface (EIDE) for specifying these workloads.}
    \label{fig:archnew}
\end{figure}

A \name{} system (Figure~\ref{fig:archnew}) consists of following components:

\begin{itemize}
\item
An \textbf{expressive integrated development environment (EIDE)} supporting a mix of programming paradigms, languages, and libraries for hardware accelerators. The EIDE facilitates application development using heterogeneous programming paradigms such as Python, SQL, Cipher, and Java including natural language interfaces. It is used by users to declare the configuration for a \name{} system such as deployment details of different data stores and architecture of hardware accelerators.

\item
A \textbf{compiler} takes in the program description from the EIDE and performs static optimizations and allocations. A compiler's frontend interfaces with the EIDE to generate an intermediate representation (IR), which is then optimized in the core, and finally the backend sends the program representation to the middleware.

\item
The \textbf{middleware}, also termed as runtime, is responsible for the actual execution of a program. As such, it consists of both the execution engine and a runtime optimizer. Similar to optimizations performed in the core of the compiler, optimizer uses cost models and optimization rules to optimize the IR for heterogeneous execution across data-processing engines and hardware accelerators. Finally, the executor coordinates execution through the adapter and data migrator (DM) using the configuration parameters. These parameters specify hardware details, such as FPGA or GPU, about the cluster data-processing engines and accelerators, their respective host CPU architectures, platform deployment details (\eg Spark, Neo4Jm, or Accumulo), and metadata (\eg location, type, and schema).

\item
An \textbf{adapter} to interface with each data-processing engine. It receives a piece of IR to transform and execute locally at the data-processing engine. Further, it collects the performance metrics after the workload execution and sends it to the middleware's optimizer.

\item
A \textbf{data migrator (DM)} manages data movement across data-processing engines in response to instructions from the middleware.
\end{itemize}

To understand the working of a \name{} system, consider a query on \textit{Admission} and \textit{Patients} tables in the MIMIC III data set~\cite{mimic}. The query requests a patient's admission history that is identified by `pid' and sorted using admission dates (`Date'). The \textit{Admission} and \textit{Patients} tables are stored in databases DB1 and DB2, respectively. DB1 projects the table \textit{Admission} on `pid,' while DB2 projects the table \textit{Patients} on `pid.' After receiving DB2 output through a data migrator, DB1 performs a sort-merge on `Date'. A \name{} system can accelerate DB1's sort operations as well as the data migration task from DB2 to DB1, pipelining it to reduce latency. 

\subsection{Acceleration in \name{} Systems}

We now discuss various acceleration opportunities in polystore systems. 

\iffalse
Hardware accelerators have limited on-chip memory. Performance of applications executed on hardware accelerators is bounded by the latency, introduced due to multiple data transfers from main memory to the accelerator memory. The accelerators are designed to exploit parallelism and pipeline behaviour of applications to gain performance benefits and mitigate the data-transfer latency.
 
A GPU is best suited for SIMD operations to get more compute per cycle, e.g. hash operation on a number of data keys is a SIMD operation. An FPGA yields comparable performance at low power and allows programming more generalized operations. FPGAs can be used as a flexible accelerator for high-cost operations within an execution plan. Due to their comparatively low memory, they're best suited for streamable operations (i.e. Filter, GroupBy on a sorted collection, etc). However, building applications on FPGAs requires a developer-friendly programming environment. CGRAs, such as plasticine~\cite{plasticine}, provide higher level abstractions to program than FPGA.
\fi

\subsubsection{Operator Execution}
The data-processing engine of a polystore system translates an application program to an intermediate representation (IR) consisting of a sequence of operators: SQL queries get mapped to projection, hash, sort, group-by, and join operators; HiveQL queries get translated to a sequence of MapReduce operations; cipher programs translate to a series of graph operations such as match, subtree, path, and join; and a deep-learning algorithms are converted into GEMV (matrix-vector multiplication) and GEMM (matrix-matrix multiplication) operations for inference and training. 

Further, the \name{} system exploits the pipeline-execution behaviour of operators to accelerate them using FPGA (\eg  bitonic sort algorithm has inherent pipeline execution~\cite{bitonic}) or execute domain-specific operators exhibiting parallel execution, such as deep-learning operators, faster using a custom hardware like TPU~\cite{tpu}.

\subsubsection{Data Access}
A data-processing engine fetches data before launching data-processing operators. Different data-processing engines support a variety of data access mechanisms depending on storage, location, size, and compression techniques. For example, relational databases employ sequential and index scans
as data-access operations, while Hadoop Distributed File System (HDFS) employs only sequential scan operations. 

A \name{} system can stream output of a sequential scan operation returning large amount of data to FPGA-based accelerator to filter and/or project relevant columns and records to reduce the amount of data communicated to the main memory for further processing.
Also, the \name{} system can cache index-seek operation(s) to reduce data-access latency.

\subsubsection{Data Migration (DM)} 
The data-processing engines in polystore systems need to have a mechanism of sending data to each other. For example, most of the data-processing engines support data migration in to comma-separated-value (CSV) format for data portability. 
So a naive approach for data migration is exporting data from the source data store to a CSV file, transferring the CSV file to the destination data store over a network, and importing the CSV file by the destination data store. This leads to multiple conversions to/from a CSV. However, Pipegen~\cite{pipegen} uses `network pipes' to eliminate disk writing and serialization while migrating data across data stores.
Pipegen transfers $10^9$ elements (4 int, 3 double), approximately 40 GB, in 35 minutes on Amazon's m4.large instances, where most of the time is spent transforming different data types into optimized binary. 

A \name{} system can offload the Pipegen's network pipes' computations and serialization algorithms to an accelerator to pipeline data transformation and network transfer. Furthermore, the system can harness Remote Data Memory Access (RDMA) accelerators to transfer data from one server's memory to another bypassing overheads of memory copy in a network protocol stack. 

\subsubsection{Adapter}
An adapter co-locates with each data-processing engine to transform and execute a piece of IR on the local data-processing engine. The transformation process of mapping operators in IR to a set of operators compliant to the local processing engine, comprises a fixed set of rules. A \name{} system can encode the data-flow graph of the rules in an accelerator to free the host CPU cycles for the local data processing.

\section{Challenges and Approaches}\label{sec:challenges}
The introduction of an accelerator in a system needs changes in all phases of application development from the programming model to its implementation. In the following sections, we discuss the new research challenges in the \name{} system stack, primarily due to introduction of hardware accelerators.

\subsection{Programming Environment (EIDE) Challenges}
\begin{figure}[t]
  \centering
  \includegraphics[width=0.65\linewidth]{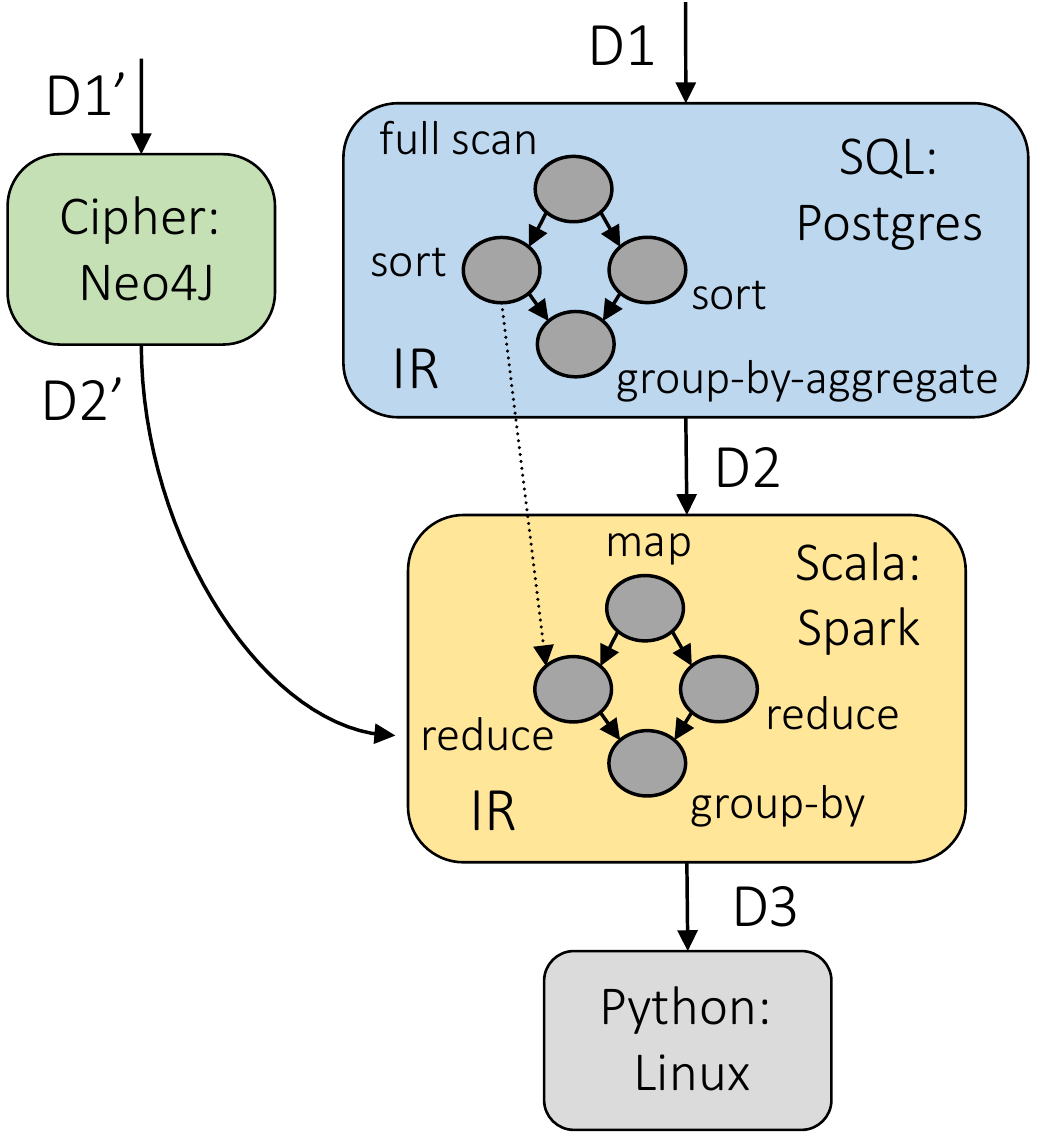}
  \caption{Heterogeneous workload abstraction as an annotated data-flow-graph. D1, D1', D2, D2' and D3 represents data sets input to each piece of source code. IR: Intermediate representation of workload in the data store. Dotted line indicates data transfer and transformation across IR nodes.}
    \label{fig:dfg}
\end{figure}

An EIDE facilitates writing clarity-optimized programs\footnote{We refer to these programs as clarity-optimized in contrast to performance-optimized. Heavy performance optimization quickly obfuscates the meaning of code, but is nearly always required to reach near-peak performance.} using well-known programming languages such as Python for ML/DL, SQL for database processing, and Cipher for graph operations. For example, Figure~\ref{fig:dfg} shows an annotated data-flow graph for a heterogeneous program. One goal of EIDE is to allow using same data across different data models such as Julia~\cite{julia} that allows a mix of C and Python data types or a multi-paradigm tensor-based model for representing heterogeneous programs. As such, EIDE's purpose is to accurately capture the computations being done and expose semantic information to the optimizer and compiler. Furthermore, an EIDE provides a clarity-optimized programming interface while simultaneously enabling efficient execution. We note that there are several key challenges in the design of such a programming environment.

\paragraph{a. What useful information can be extracted from each subprogram to execute them optimally} Heterogeneous programs by definition consist of a mix of programming paradigms and abstraction levels. In order to capture and later optimize these programs, such a framework should be able to losslessly capture semantic information from each sub-program. This information is likely to be used in scheduling, resource allocation, as well as operation selection. For example, information about numerical precision may be used to guide between different matrix multiplication algorithms.

\paragraph{b. How to transform same data across different data models} Models and storage formats differ across programming domains, and overheads incurred by data movement and transformation across domains can quickly exceed benefits of acceleration. The related work in the area of communication-avoiding algorithms indicates that data movement can easily become a bottleneck across a distributed system if not properly optimized~\cite{Demmel2008}. For example, a large model such as the Google Neural Machine Translation~\cite{GNMT} model may ordinarily have a few gigabytes of weights when stored efficiently. If this data were stored in a textual format, the weight's storage cost balloons into the terabyte range. Such an increase in required data movement will significantly impact the performance of the system as a whole.

\paragraph{c. How to design a multi-paradigm language} It appears unlikely that any particular sets of domain-specific languages and base languages will suffice to cover computing needs in the long term. As a consequence, EIDE must be able to accommodate novel languages and abstraction levels.

\paragraph{d. What functions should be accelerated} 
Trivially, everything that can be accelerated should be accelerated, but in practice identifying such opportunities can be difficult. With  reconfigurable hardware, nearly everything can be accelerated to varying degrees of profitability; as a result, a \name{} system needs to solve the additional problem of area and bandwidth allocation on these accelerators. Additionally, performance characterizations on accelerators such as FPGAs tend to be inaccurate without repeated synthesis, a process which takes hours to days per run for non-trivial designs.

\paragraph{e. How should a natural language query be compiled to a semantic equivalent heterogeneous program} 
Yaghmazadeh et al.~\cite{sqlizer} and Jagdish et al.~\cite{sqlthesis} have approached the problem of translating a natural-language query to a SQL query. Virtual assistants such as Almond~\cite{almond} convert natural language commands into programs. These assistants may be extended to incorporate work such as AutoML~\cite{automl} to automatically generate machine learning models for more complex queries.

\subsection{Compilation Challenges}
We can separate the structure of a traditional compiler into three components: frontend, core, and backend. In a heterogeneous system, their tasks are adapted for inter-operation of both frontend and backend components. The Intermediate Representation (IR) is a data structure which represents the program as a whole.

\subsubsection{IR Design}
One of the goal in designing IR is to capture semantics of data processing on different engines and hardware accelerators and should be extensible to incorporate semantics of new compute engines.

\begin{figure}[t]
  \centering
  \includegraphics[width=0.95\linewidth]{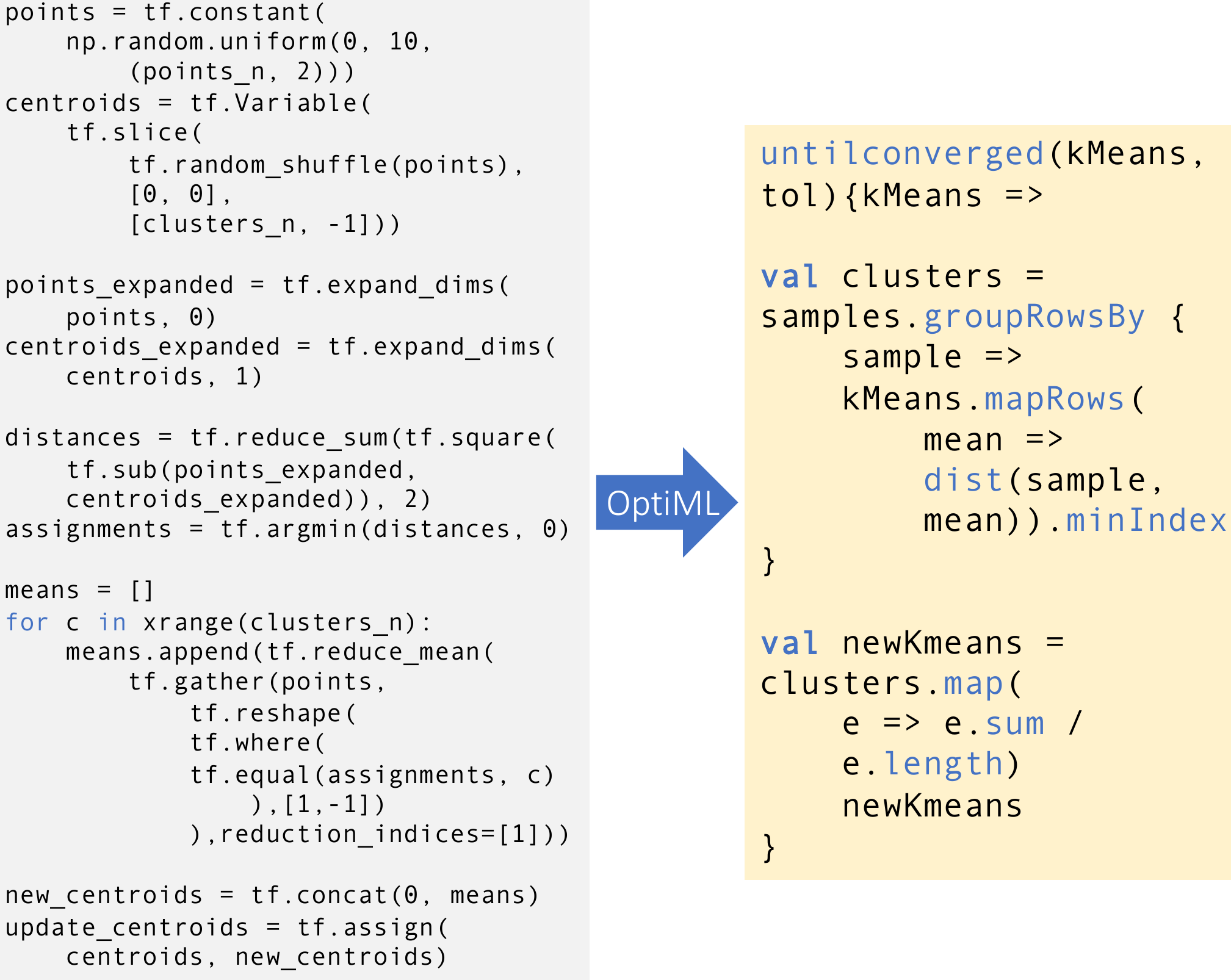}
  \caption{An example of translating a Tensorflow K-Means application to OptiML~\cite{optiml}, a domain-specific language for hardware accelerators.}
    \label{fig:optiml}
\end{figure}

A program on data-processing engines is more like a control-flow graph, whereas a hardware accelerator works on data-flow graph. One approach is to have a hierarchical IR consisting of control nodes and each control node may have a data-flow graph for an operator as shown in Figure~\ref{fig:dfg}. 
The program inside a control node may get converted into a set of available hardware domain-specific operators for generating target specific code. 
Figure~\ref{fig:optiml} shows an OptiML example of short k-means application translated from longer Tensorflow version, that can be optimized for multi-core CPU, GPU or FPGA accelerators~\cite{spatial,delite}.

\begin{figure}[t]
  \centering
  \includegraphics[width=\linewidth]{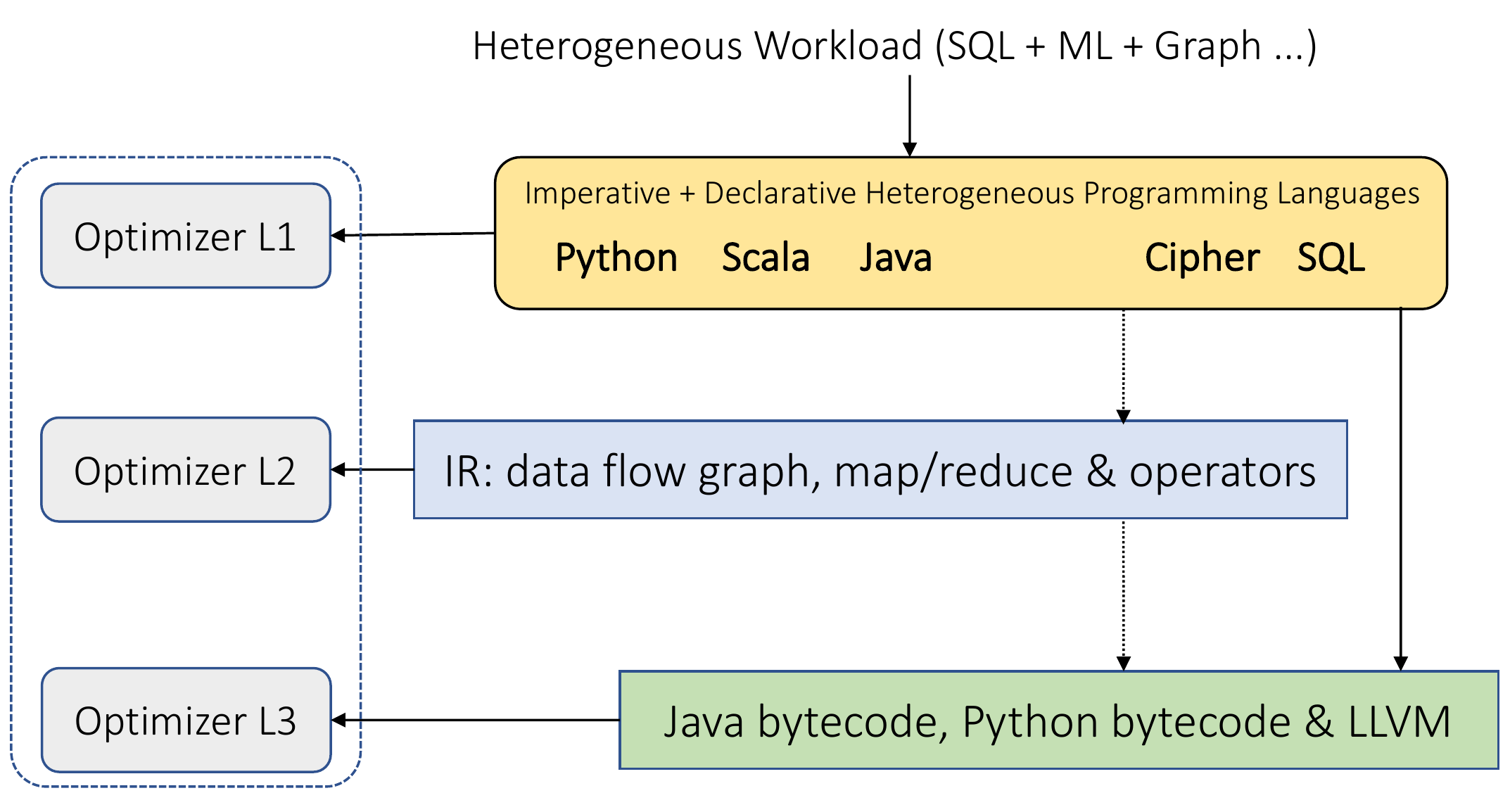}
  \caption{Levels of optimization for a heterogeneous workload -- L1: piece of program into DSL, L2 and L3: IR and lowest level of source code respectively at individual data-processing engines.}
    \label{fig:opt}
\end{figure}

\subsubsection{Frontend}
A typical compiler frontend consists of a parser, followed by a variety of checks. These may include type checking and static assertions, as well as elaborating the code. For example, the C frontend handles macro expansions and \textit{include}s. In a heterogeneous compiler for \name{} systems, the frontend faces the task of constructing a compute graph from a variety of sub-programs. Depending on the construction of the compiler, the frontend must still perform inter-subprogram checks, but can delegate subprogram-local checks to their respective frontends.

\subsubsection{Core}
The core of a compiler is where most interesting transformations and optimizations are made. Ideally, such a core can reuse optimizations across a variety of applications. In order to do so, it must be able to express these optimizations in a domain-agnostic manner in \name{} environment. Unlike lower-level optimizations such as loop transforms, and strength reductions, these high-level optimizations may require semantic information. These high-level transformations cover L1 optimizations in Figure \ref{fig:opt}. For example, when computing the eigenvalues of a matrix either Lanczos or Householder tri-diagonalization methods may be used. These algorithms create different resultant matrices, but both are valid for computing eigenvalues.

While instruction selection and scheduling are part of a traditional backend's responsibilities, they must be moved to the core due to their interaction with resource and task allocation. In a \name{} system, the core must decide where each task should be assigned, and which resources may be attached (\ie storage or network bandwidth). Resource allocation is particularly important as a transformation's benefits
are intimately tied with the characteristics of the executing platform.

\subsubsection{Backend}
The role of the backend in a heterogeneous system is the least changed part compared to a traditional compiler. In a heterogeneous system the backend is responsible for generating the code corresponding to each computing unit, involved. Each constituent computing  unit is then responsible for providing component-specific optimizations, such as fine-grained tiling, pipelining, and other local tasks in FPGA. In a \name{} system, the high-level decisions made by different constituent components become L2 optimizations in Figure \ref{fig:opt}. Finally, L3 optimizations are the implementation-level transformations for each component, such as those in the underlying C/C++ compiler for a database engine. We note here that while the Roofline Model \cite{roofline} is used as a method of measuring performance on CPUs and can be extended to fixed hardware, however, obtaining accurate and simple analytical models of theoretical performance on reconfigurable hardware such as FPGAs or CGRAs is substantially more difficult. One efficient technique can be found in Koeplinger et. al. \cite{koeplinger2016} which uses repeated sampling to build an empirical performance model.

\subsection{Optimization Challenges}
Optimization challenges arise in both the middleware and compiler. In the middleware these tasks are delegated to the optimizer, which minimizes the total execution time of a program, while optimizing on number and size of data movements and cost of operators' execution across data stores~\cite{rheemopt}. In the compiler for a \name{} system, such parameter selection is termed tuning or auto-tuning, and naturally arises from parallelism factors, numerical precision, and resource allocation.

Mathematically, in the mono-objective formulation, this is the problem of finding a global minimizer of an unknown (black-box) objective function $f$:
\begin{equation}\label{eq:optim_pb}
	x^* = \argmin_{x \in \bbX} f(x)
\end{equation}
where $\bbX$ is an input design space of a \name{}, which includes configuration of heterogeneous computing units and design parameters of hardware accelerators.
The problem addressed in this paper is the optimization of a deterministic
%function $f:\bbX \rightarrow \bbR^m$ over a 
black-box function $f:\bbX \rightarrow \bbR$, \eg a computer program, over a domain of interest that includes lower and upper bound constraints on the problem variables.
The design space of the optimization in \name{} includes discrete variables, \ie either categorical (\eg boolean) or ordinal (\eg choice of on-chip memory sizes); 
these type of variables preclude the computation of derivatives (derivatives are not defined on discrete variables) and by consequence the use of standard gradient-based optimization procedures.
Hence, we assume in \name{} that the derivative of $f$ is neither symbolically nor numerically available.
This problem is referred to in the mathematical optimization literature as black-box optimization \cite{feliot2017bayesian} and, in the computer systems community as design space exploration (DSE) \cite{ipek2006efficiently,kang2010approach}.

%A similar example is in creating hardware designs for field-programmable gate arrays (FPGAs). FPGAs are a type of reconfigurable logic chip with a fixed number of units available to implement circuits. 
%For example, any generated design for a hardware accelerator, such as an FPGA, must keep the number of units strictly below available resource budget to be implementable on the FPGA. 
%In these examples, feasibility of an experiment cannot be checked prior to termination of the experiment; this is often referred to as \emph{unknown feasibility} in the literature \cite{gelbart2014bayesian}.
HyperMapper 2.0~\cite{Hypermapper20,bodin2016integrating,nardi2017algorithmic} is designed for heterogeneous workloads, and  can handle complex design spaces consisting of multiple objectives, categorical/ordinal variables, unknown feasibility constraints \cite{gelbart2014bayesian}, and exploitation of performance profiling of earlier executions of workloads in \name{}.

\subsubsection{Multi-Objective Optimization}
\label{active_learning}
One approach to achieve multi-objective optimization is to build cost models for each of the heterogeneous computing units. 
A cost model could be a performance prediction model, also known as surrogate models, to estimate performance efficiency of workload on different platforms for scheduling, \eg Luo~et al.\cite{queryeff} presents power-performance characteristic of analytic queries on database systems, and Singhal et al.~\cite{SQLperf},~\cite{hadoopperf},~\cite{sparkperf} and ~\cite{hiveperf} present similar cost models for relational database, Hive, Hadoop, and Spark platforms deployed on conventional CPU systems. However, none of these include design space for hardware accelerators. 

%One goal of optimization is to build new cost models for operators, perhaps using these performance prediction models where cost is a function of input data size, a data-processing engine, a hardware accelerator, hardware details.
The input data size could be a number of items, a data-processing engine could be Spark, Hadoop, Postgres, or more; hardware accelerator could be FPGA, GPU, CPU-MPI; and hardware details in terms of memory size, FPGA board LUT units or CGRA layout or GPU memory with the connector. 

The optimizer has large design space to explore for a \name{} system. Moreover, building cost models with a large number of parameters may be expensive. Alternatively, active learning~\cite{bodin2016integrating},~\cite{nardi2017algorithmic},~\cite{Hypermapper20} can trade off exploration and exploitation mechanisms to give an approximated optimal configuration for workload execution in \name{} systems.
%EXPLORE What role should ML play in the optimization  process. Can we use active learning to search the space ??????
%The cost model shall estimate execution time of a function execution on accelerator as function of the input data size.  The cost model can be updated by collecting performance numbers from adapter of data-processing engines. In addition, the Optimizer also need to optimize the layout of the kernel on the hardware for FPGA/CGRA. 

Active learning is a paradigm in supervised machine learning which uses fewer training examples to achieve better optimization by iteratively training a predictor, and using the predictor in each iteration to 
choose the training examples which will increase its chances of finding better configurations and at the same time improving the accuracy of the prediction model. 
Thus the optimization results are incrementally improved by interleaving exploration and exploitation steps. 
One can use randomized decision forests~\cite{breiman2001random} as the base predictors created from a number of sampled points in the parameter space, which is different configurations of the \name{} design space.

\begin{figure}[t]
  \centering
  \includegraphics[width=\linewidth]{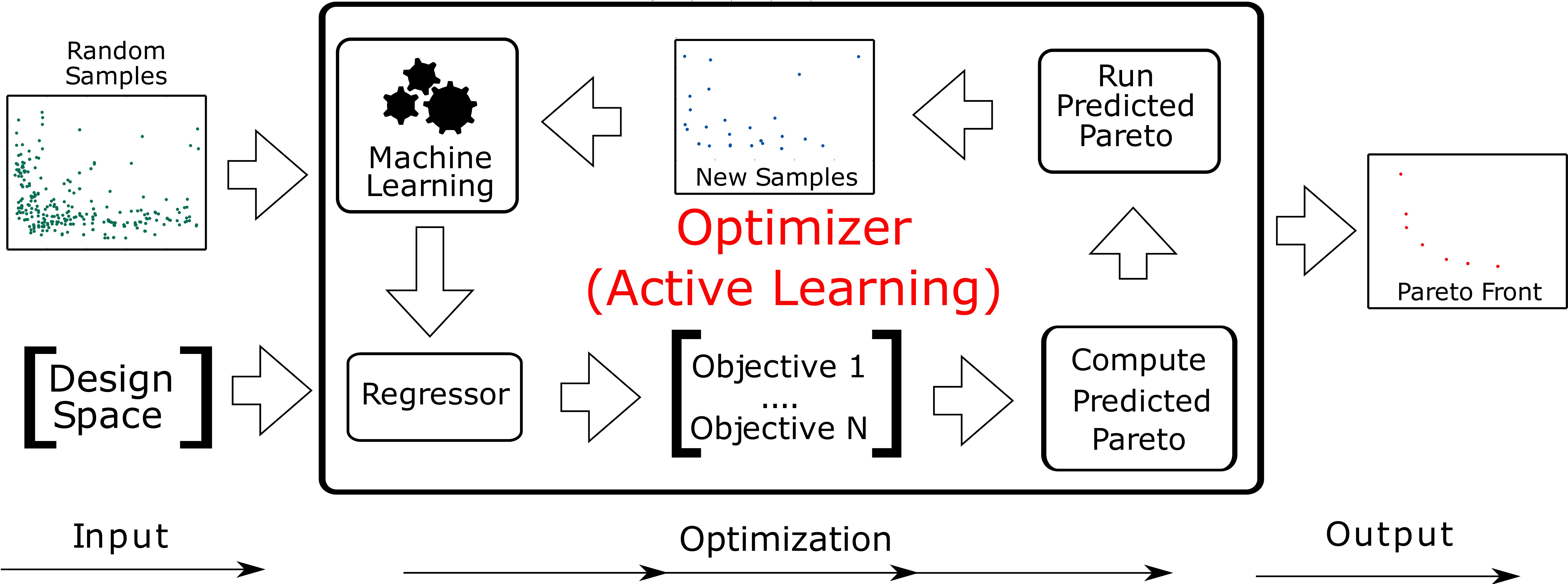}
  \caption{Active learning is a more efficient method of searching the design space by selecting evaluation points which are likely to reveal information about good configurations. In a multi-objective setting, the system attempts to learn the Pareto curve, a generalized notion of optimality.}
    \label{figure_active_learning}
\end{figure}

The application is evaluated on the sampled points, yielding the labels of the supervised setting given by the multiple objectives which could be obtained from the historical executions of workloads on \name{}.
Since our goal is to accurately estimate the points near the Pareto optimal front, we use the current predictor to provide performance values 
over the parameter space and thus estimate the Pareto fronts. 
For the next iteration, 
only parameter points near the predicted Pareto front are sampled and evaluated, and subsequently used to train 
new predictors using the entire collection of training points from current and all previous iterations. The evaluation of sample point means actual execution of the workload in the given configuration, which may execute sub-optimally, however, it helps learning the cost model for future similar workloads. This process is repeated over a number of iterations forming the active learning loop. Bodin et. al. ~\cite{bodin2016integrating} and Nardi et. al.~\cite{nardi2017algorithmic} indicate that this guided method of searching for highly informative parameter points in fact yields superior predictors as compared to a baseline that uses randomly sampled points alone.
By iterating this process several times in the active learning loop, we are able to discover high-quality design configurations that lead to good performance outcomes. Figure \ref{figure_active_learning} shows the high-level active learning DSE algorithm.

%\begin{center}
%\begin{figure}[tb]
%\includegraphics[scale=0.13]{images/active_learning_revisited.pdf}
%\caption{Active learning optimizer.}
%\label{figure_active_learning}
%\end{figure}
%\end{center}

\subsection{Execution Challenges}
The goal of the executor is to schedule the optimized IR and coordinates its execution across heterogeneous computing units. The optimized IR may be considered to be a sequence of stages (like Spark), where each stage may have heterogeneous tasks to execute in parallel and a task could be an operator on a data-processing engine or hardware accelerator or a data movement. The whole workload execution can be perceived as a pipeline of the stages' execution for maximum throughput. An operator may be implemented as a stream, where data may arrive from local storage or remote platform through the accelerator. 
%Kernels for hardware accelerators can be converted to accelerator specific code using Spatial and Delite compilers. The executor interacts with Data migrator(DM) and adapter for the IR execution. 

The research challenges for the execution are as follows:

\paragraph{a. IR mapping to local accelerators and kernels} At runtime a variety of candidates may be available, and the selection will ultimately depend on a combination of the runtime environment and data-dependent analyses.

\paragraph{b. Coordination and scheduling across heterogeneous units} The field of scheduling is well studied, but is a difficult engineering challenge, especially due to the complexity of interfacing with accelerators.

\paragraph{c. Runtime acceleration} Hardware acceleration motivates design and development of data flow algorithms for the adapter and DM for a set of data-processing engines. Customized hardware can also be explored for data migration acceleration since it is a prevalent operation in data centers.

\paragraph{d. Runtime statistics}
Runtime metrics are crucial for optimization, but hardware accelerators do not provide such utilities. Currently, such metrics may be instead gathered through low-level simulators, but run orders of magnitude slower than actual hardware.

\section{Conclusion}\label{sec:concl}
Modern real-time analytic applications are represented using a mix of heterogeneous programming paradigms (referred to as heterogeneous workload) spanning across heterogeneous data-processing engines working with different data models. Polystore systems suit such applications, but with the decline of Moore's Law and Dennard Scaling, current and future workloads will require hardware accelerators to meet performance requirements. 

%We envisioned a next-generation polystore systems, referred to as \name{}, using hardware accelerators to achieve performance at low power. We have presented an architecture for \name{} system with acceleration for data-processing engines, their adapters, and the data migrator across these engines.

We present \name{}, an architecture to accelerate polystore systems using specialized hardware to achieve performance at low power. This architecture highlights various open and novel challenges, which may guide future work in the area and provide end-to-end performance and power improvements. We note that these challenges span beyond just the database world, but also encompass the compiler, optimization, and hardware communities.

\section*{Acknowledgments}
\addcontentsline{toc}{section}{Acknowledgment}
The authors wish to thank Jeffrey D. Ullman and Tian Zhao for helpful early discussions; and the anonymous ICDCS reviewers for their feedback on earlier versions of this paper. This work is based on research sponsored by Air Force Research Laboratory (AFRL) and Defense Advanced Research Projects Agency (DARPA) under agreement number FA8650-18-2-7865. The U.S. Government is authorized to reproduce and distribute reprints for Governmental purposes notwithstanding any copyright notation thereon. The views and conclusions contained herein are those of the authors and should not be interpreted as necessarily representing the official policies or endorsements, either expressed or implied, of Air Force Research Laboratory (AFRL) and Defense Advanced Research Projects Agency (DARPA) or the U.S. Government. This research is also supported in part by Tata Consultancy Services, affiliate members and supporters of the Stanford DAWN project: Ant Financial, Facebook, Google, Infosys, Intel, Microsoft, NEC, Teradata, SAP, and VMware.

%\bibliographystyle{./IEEEtran}
%\bibliography{./IEEEabrv,./IEEEexample}

\bibliographystyle{IEEEtran}
\bibliography{icdcs}

\end{document}